\documentstyle[preprint,aps,epsf]{revtex}
\newcommand{\be}{\begin{equation}}
\newcommand{\ba}{\begin{eqnarray}}
\newcommand{\ee}{\end{equation}}
\newcommand{\ea}{\end{eqnarray}}
\newcommand{\nn}{\nonumber}

\newcommand{\e}{\mbox{e}}
\newcommand{\sla}[1]{#1 \!\!\! \slash}
\renewcommand{\theequation}{\arabic{section}.\arabic{equation}}
\begin{document}
\draft
\preprint{DESY 97-087, hep-ph/9705362}
\date{May 1997}
\title{On the Possibilities of Distinguishing\\ 
       Majorana from Dirac Neutrinos}
\author{Axel Hoefer\footnote{electronic address: ahoefer@ifh.de}}
\address{DESY--IfH Zeuthen, Platanenallee 6 \\
D-15738 Zeuthen, Germany}
\maketitle
\begin{abstract}
Cross sections involving massive Dirac or Majorana neutrinos usually
differ only by terms which are suppressed by the smallness of the
neutrino mass. The process $e^+e^-\to\nu\nu$, 
however, is an example
in which the two differential cross sections are quite distinct even in
the limit $m_\nu\to 0$ as long as $m_\nu\not=0$. I discuss the cause of
this phenomenon and comment on strategies to identify processes with
quite different Majorana and Dirac neutrino signals.
\end{abstract}
\newpage
\section{Introduction}
The question of distinguishing Dirac from Majorana neutrinos has been
frequently discussed (see e.g.~\cite{moha,gibrat}).  Since the
Majorana neutrino is its own anti-particle, it has in general quite
distinct properties from the Dirac neutrino.  Experimentally, it is
nevertheless difficult to distinguish between these two kinds of
neutrinos.  I discuss the possibility of revealing the neutrino's
nature using neutral-current (NC) neutrino data such as measured at
LEP I or by the CHARM II collaboration.  Because of contradicting
statements which have recently appeared\cite{plaga,kayser-c,hanne} I
provide explicit results for the neutrino cross sections of concern,
keeping the neutrino mass dependence where necessary. In particular I
illustrate and comment on the following list of important and more or
less familiar properties of neutrinos:

\begin{description} 
\item[a)] If neutrinos are massless particles {\it and} if only the
  left-handed (or, equivalently, only the right-handed) field
  interacts then Majorana and Dirac neutrinos are identical, such that
  the theoretical distinction into ``Dirac'' and ``Majorana'' is
  meaningless.\cite{shrock} This is particularly true for standard model (SM)
  neutrinos. The proof is given in Appendix~A.
\item[b)] A massless neutrino ($m_\nu=0$) is always a helicity
  eigenstate for {\it both} 
  Majorana and Dirac neutrinos. In this case the
  helicity is an intrinsic property and represents a Lorentz
  invariant quantum number.
\item[c)] A {\it massive} Majorana neutrino, even in the small-mass limit
  ($m_{\nu}\to 0$ with $m_\nu\not=0$), is in general not an eigenstate
  of helicity and can have a spin pointing in an arbitrary direction. 
  This is also true for massive Majorana
  neutrinos moving at relativistic velocities ($\beta\to 1$,
  $\beta\not=1$); see e.g.~\cite{hoefer,Ma}.
\item[d)] As a consequence of b) and c), observable quantities for 
  Majorana neutrinos with small non-vanishing masses
  are in general different from those of 
  massless Majorana neutrinos,
  and they remain distinct even in the limit of $m_\nu\to0$, $m_\nu\not=0$. 
  Similarly, massless Majorana neutrinos and highly
  relativistic massive Majorana neutrinos have in general
  distinguishable  observables even in the limit $\beta\to1$,
  $\beta\not=1$. In other words, the limits $m_\nu\to0$ and
  $\beta\to1$ are {\it not} smooth for Majorana neutrinos.\cite{boudjema} 
  An example
  of this important and often neglected point are neutrino 
  cross sections for LEP I which are given in Sec.~II.
\item[e)] An exception of d) are processes in which neutrinos are
  produced via a weak charged current (neglecting here
  the possibility of CC neutrino {\it pair} production):
  If the charged current (CC) is
  purely left-handed (or, equivalently, purely right-handed)
  observables of these processes are approximately the same for {\it
    light} ($m_{\nu}\to 0$ with $m_{\nu}\not=0$) and for {\it
    massless} ($m_{\nu}=0$) neutrinos, disregarding whether the
  neutrinos are Majorana or Dirac particles.\footnote{ In these cases
    both light and massless neutrinos are helicity eigenstates.} The
  limit $m_\nu\to0$ is smooth.  This feature is illustrated by an
  explicit calculation of neutrino cross sections for the CHARM II
  experiment; see Sec.~III and Appendix B.
\item[f)] A massive Dirac neutrino in the small-mass limit
  ($m_{\nu}\to 0$ with $m_\nu\not=0$) is approximately an eigenstate
  of helicity.  Unlike in the case of Majorana neutrinos, the limit
  $m_\nu\to0$ is always smooth: Physical observables calculated for
  massive Dirac neutrinos are identical to those of massless Dirac
  neutrinos when taking $m_\nu\to0$.  (Similarly, the relativistic
  limit $\beta\to 1$ is smooth.) This property of Dirac neutrino cross
  sections is apparent in the results given in Sec.~II and III.
\item[g)] The neutral current coupling of a massive Majorana neutrino
  is pure axial-vector (assuming diagonal NCs), independent of the
  specific values of the vector and axial-vector couplings in the
  Lagrangian.  This means that the vector coupling of a massive
  Majorana neutrino does {\it not} contribute to the neutral current,
  no matter what its value is.  This follows from charge-conjugation
  properties of the free Majorana field; see Appendix A.  In contrast,
  the vector coupling of a (massless or massive) Dirac neutrino {\it
    does} contribute to the neutral current, such that the general
  neutral current cross sections of Dirac neutrinos consist of both
  vector and axial-vector contributions.
\end{description}

In Sec.~IV, I provide a summary, focusing on the statements d) and e)
given above. I point out the importance of the history of an incoming
neutrino, in particular, whether it has been produced in a (chiral)
charged current interaction or not.  Furthermore, I comment on
previous publications on the subject of distinguishing Majorana and
Dirac neutrinos.  The possibility of flipping the helicity
of the neutrino through strong magnetic fields is also considered.
\section{Neutrino cross sections at LEP I}
\setcounter{equation}{0} 
Neutral current neutrino pair production has been measured indirectly
at LEP I.  Taking the neutrino couplings to be the purely
left-handed\footnote{For general neutrino couplings, see Appendix B.}
neutrino interaction of the SM Lagrangian, the angular distribution
for $e^+e^-\to\nu_f\bar\nu_f$ can be easily calculated.  At the Z peak
[$s\approx m_Z^2$, $s$ being the squared center-of-mass-system (CMS)
energy], the neutrinos are (almost\footnote{In the case of $f=e$, a
  charged current contribution exists, but can be neglected for
  $s\approx m_Z^2$.}) exclusively produced via neutral current
exchange (Fig.~1).  In the case of Majorana (M) neutrinos
($\bar\nu\equiv\nu$) and Dirac (D) neutrinos the differential cross
sections at $s\approx m_Z^2$ are
\ba
\left(\frac{d\sigma}{d\Omega}\right)_{M}^{m_\nu\not=0}&=&\sigma_0\;\biggl\{
\left[(g_V^e)^2+(g_A^e)^2\right](1+\cos^2\theta)\;
+O\left(\frac{m_e^2}{m_Z^2}\right)
+O\left(\frac{m_\nu^2}{m_Z^2}\right)\biggr\}\;,\\ 
\left(\frac{d\sigma}{d\Omega}\right)_{D}\quad\,\;&=&\sigma_0\;\biggl\{
\left[(g_V^e)^2+(g_A^e)^2\right](1+\cos^2\theta)\;+\;4g_V^eg_A^e\cos\theta
\;\nn\\ 
&&\qquad+O\left(\frac{m_e^2}{m_Z^2}\right)+O\left(\frac{m_\nu^2}{m_Z^2}\right)
\biggr\}\;,
\ea 
with 
\ba 
\sigma_0 &=& \frac{G_F^2s}{128\pi^2}\;
\left|\frac{m_Z^2}{s-m_Z^2+im_Z\Gamma_Z}\right|^2\;, 
\ea 
$\theta$ being the CMS angle between the momenta of the initial
electron and the outgoing neutrino, and $g_V^e$ ($g_A^e$) being the
vector (axial-vector) coupling of the electron to the Z boson.
Setting $m_\nu=0$, the Dirac differential cross section is identical
to the SM result for massless neutrinos.  The Majorana and Dirac
differential cross sections, however, {\it remain different even in
  the limit $m_\nu\to0$, $m_\nu\not=0$}. This
illustrates the 
statement d) of the introduction.  Setting $m_\nu=0$, (2.1) is no
longer valid because massless and massive Majorana neutrinos are
completely different objects; see Appendix~A.  The cross section of a
massless Majorana neutrino is actually equal to the massless Dirac
neutrino cross section and is obtained by taking the limit $m_\nu\to0$
in (2.2). In the case of generalized couplings, the massless Majorana
and Dirac cross sections are not identical; see Appendix~B.

Measuring the angular distribution of the neutrino pairs would clearly
provide for a conclusive answer on the Dirac or Majorana nature of
massive neutrinos.  In particular, the forward-backward 
asymmetries of
(2.1) and (2.2) are
\ba
A_M^{FB}&=&0\;,\\
A_D^{FB}&=&\frac{3}{2}\;\frac{g_V^eg_A^e}{(g_V^e)^2+(g_A^e)^2}
\approx0.45
\ea
for Majorana and Dirac neutrinos, respectively. In a somewhat
different context such asymmetries have already been considered in
\cite{hoefer,bilenky,petcov}. 
Even if the CC contribution to neutrino pair production was included,
$A_M^{FB}$ would remain zero and would still be 
significantly distinct from $A_D^{FB}$. Hence measuring the
differential cross section for neutrino pair production 
would determine the nature of the neutrino even at CMS 
energies far away from the Z peak.

Unfortunately, the
measurement of the angular distribution of the outgoing neutrino-pair
is not feasible at LEP I since the number of produced neutrino pairs
(about 2 million) is far too small to create a sufficient number of
detected neutrino-pair events in the LEP experiments.

Assuming lepton flavour universality and the existence of three light
neutrino generations, the LEP I data on the Z peak provide for an {\it
  indirect} measurement of the {\it total} cross section for NC
neutrino-pair production.  Integrating (2.1) and (2.2), the
theoretical predictions of the Dirac and Majorana total cross sections
are found to be identical up to terms of $O(m_\nu^2/m_Z^2)$. [The
terms of $O(m_e^2/m_Z^2)$ in (2.1) and (2.2) are the same.]  Taking
into account the existing experimental upper bounds on light neutrino
masses\cite{partdata}, the $O(m_\nu^2/m_Z^2)$ difference in the total
cross sections is too small to be tested by the LEP I experiments.
Hence LEP I is not able to distinguish between light Dirac and light
Majorana neutrinos.


\section{Neutrino cross sections of the CHARM II Experiment}
\setcounter{equation}{0} 
Neutral current neutrino cross sections have also been measured by the
CHARM II collaboration\cite{vilain} using the processes $\nu_\mu
e^-\to \nu_\mu e^-$ and $\bar\nu_\mu e^-\to \bar\nu_\mu e^-$
\cite{hooft,sehgal}. The
incident neutrinos are produced via a chiral CC interaction.  
They are therefore approximately eigenstates of helicity.
This needs to be taken into account when calculating the relevant cross
sections: Recall point e) of the introduction.

The CHARM II experiment proceeds as follows: Charged pions are produced and
directed at a fixed target.  They predominantly decay into a muon and
a muon neutrino.  This neutrino can interact with a target electron 
only via neutral current Z exchange, leading to an electron and a muon
neutrino in the final state. 

In the case of Dirac neutrinos, a positively (negatively) charged pion
can decay into $\mu^+{\nu}_{\mu}$ ($\mu^-\bar{\nu}_{\mu}$). The cross
sections for neutrino and anti-neutrino scattering are different.

In the case of Majorana neutrinos there is no distinction between
neutrino and anti-neutrino. However, the fact that the intermediate 
neutrino has been produced via a chiral CC interaction is important
and affects the subsequent neutral current interaction.
To take this into account, I calculate the complete process
\ba
\pi^+(k_1)&\to&\mu^+(k_3)+\nu_{\mu}(q)\nn
\ea
with the subsequent NC reaction
\ba
\label{subreact}
\nu_{\mu}(q)+e^-(p_1)&\to&\nu_{\mu}(k_2)+\e^-(p_2)\;.
\ea
The corresponding Feynman diagram is shown in Fig.~2.
The intermediate neutrino $\nu_{\mu}(q)$ is a virtual particle.
It appears as a propagator in the amplitude.
For simplicity I assume $\nu_{\mu}$ to be approximately a mass
eigenstate (small mixing angles in the neutrino sector).
The $2\to3$ amplitudes involving a Dirac (D) or Majorana (M) muon
neutrino are obtained as
\ba
{\cal M}_{D}^{\mu^+}&=&A\; j_{\lambda}^e\;\bar{u}_{t_2}(k_2)
               \;\gamma^{\lambda}\;P_L\;(\sla{q}+m_{\nu})\;
               \gamma_{\rho}\;P_L\; v_{t_3}(k_3)\cdot
               k_1^{\rho}\;,\\
{\cal M}_{M}^{\mu^+}&=&-A\; j_{\lambda}^e\;\bar{u}_{t_2}(k_2)
               \;\gamma^{\lambda}\;\gamma_5\;(\sla{q}+m_{\nu})\;
               \gamma_{\rho}\;P_L\; v_{t_3}(k_3)\cdot
               k_1^{\rho}\;,
\ea
where the superscript $\mu^+$ indicates the detection of a positively
charged muon in the final state.  In the case of an incoming
negatively charged pion, the three-particle final state contains a
$\mu^-$ and the amplitudes involving a Dirac or Majorana neutrino are
\ba
{\cal M}_{D}^{\mu^-}&=&A\; j_{\lambda}^e\;\bar{u}_{t_3}(k_3)\;
               \gamma_{\rho}\;
               P_L\;(\sla{q}+m_{\nu})\;
               \gamma^{\lambda}\;P_L\; v_{t_2}(k_2)\;
               k_1^{\rho}\;,\\
{\cal M}_{M}^{\mu^-}&=&-A\; j_{\lambda}^e\;\bar{u}_{t_3}(k_3)\;
               \gamma_{\rho}\;
               P_L\;(\sla{q}+m_{\nu})\;
               \gamma^{\lambda}\;\gamma_5\; v_{t_2}(k_2)\cdot
               k_1^{\rho}\;.
\ea
Here the following abbreviations have been used:
\ba
\quad q&=&k_1-k_3\;,\nn\\
j_{\lambda}^e &=& \bar{u}_{s_2}(p_2)\;\gamma_{\lambda}\;(g_V^e-g_A^e
\gamma_5)\;u_{s_1}(p_1)\;,\nn\\
P_{L}&=&\frac{1}{2}(1-\gamma_5)\;,\quad\;P_{R}=\frac{1}{2}(1+\gamma_5)\;,\\
A &=& \biggl(\frac{g}{2\cos\theta_W}\biggr)^2\;\frac{g^2}{8m_W^2}
       \;\frac{1}{(p_2-p_1)^2-m_Z^2}\;\frac{1}{q^2-m_{\nu}^2}
       \; f_{\pi}\cos\theta_1\;,\nn
\ea
$f_{\pi}$ being the pion decay constant, and $\theta_1$ 
being the CKM mixing angle for $\pi^{\pm}$
decay\cite{commins}.
 
The axial-vector coupling of the Majorana neutrino to the $Z$ boson  
is apparent in (3.3) and (3.5).
Splitting the amplitudes involving a Majorana neutrino 
into a part with  left-handed
and a part with  right-handed $Z\nu\nu$ vertex,
they can be written as ($\gamma_5=P_R-P_L$)
\ba          
{\cal M}_{M}^{\mu^{\pm}}&=&{\cal M}^{\mu^{\pm}}_{D}+
\Delta{\cal M}^{\mu^{\pm}}\;\;,
\ea
where I define
\ba
\Delta{\cal M}^{\mu^+}&=&-A\; j_{\lambda}^e\;\bar{u}_{t_2}(k_2)\;
               \gamma^{\lambda}\;P_R\;(\sla{q}+m_{\nu})\;
               \gamma_{\rho}\;P_L\; v_{t_3}(k_3)\;
               k_1^{\rho}\;,\\
\Delta{\cal M}^{\mu^-} &=&-A\;
j_{\lambda}^e\;\bar{u}_{t_3}(k_3)\;\gamma_{\rho}\;P_L\;(\sla{q}+m_{\nu})\;
               \gamma^{\lambda}\;P_R\; v_{t_2}(k_2)\;
               k_1^{\rho}\;.
\ea
In the case of general vector and axial-vector couplings, it is {\it also}
possible to write the resulting Dirac and Majorana amplitudes 
as a combination of the above results for ${\cal M}^{\mu^{\pm}}_{D}$ and  
$\Delta{\cal M}^{\mu^{\pm}}$, the coefficients being functions of the
generalized coupling constants; see Appendix B. 

Using the identities
\ba
k_1\;=\;k_3+q\;,\qquad\sla{q}\sla{q}=q^2\;,\qquad
\sla{k}_3\;v_{t_3}(k_3)\;=\;-m_{\mu}\;v_{t_3}(k_3)\;\;,
\ea
I arrive at the following expressions:
\ba
{\cal M}_{D}^{\mu^+}&=&A\; j_{\lambda}^e\;\biggl(\;q^2\;
\bar{u}_{t_2}(k_2)\;\gamma^{\lambda}\;P_L\;v_{t_3}(k_3)\;-\;
m_{\mu}\;\bar{u}_{t_2}(k_2)\;\gamma^{\lambda}\;P_L\;\sla{q}\;v_{t_3}(k_3)\;
\biggr)\;,\\
\Delta{\cal M}^{\mu^+}&=&A\; j_{\lambda}^e\; m_{\nu}\;\biggl(\;
m_{\mu}\;\bar{u}_{t_2}(k_2)\;\gamma^{\lambda}\;P_R\;v_{t_3}(k_3)\;-\;
\bar{u}_{t_2}(k_2)\;\gamma^{\lambda}\;P_R\;\sla{q}\;v_{t_3}(k_3)\;
\biggr)\;,\\
{\cal M}_{D}^{\mu^-}&=&A\; j_{\lambda}^e\;\biggl(\;q^2\;
\bar{u}_{t_3}(k_3)\;\gamma^{\lambda}\;P_L\;v_{t_2}(k_2)\;+\;
m_{\mu}\;\bar{u}_{t_3}(k_3)\;\sla{q}\;\gamma^{\lambda}\;P_L\;v_{t_2}(k_2)\;
\biggr)\;,\\
\Delta{\cal M}^{\mu^-}&=&-A\; j_{\lambda}^e\; m_{\nu}\;\biggl(\;
m_{\mu}\;\bar{u}_{t_3}(k_3)\;\gamma^{\lambda}\;P_R\;v_{t_2}(k_2)\;+\;
\bar{u}_{t_3}(k_3)\;\sla{q}\;\gamma^{\lambda}\;P_R\;v_{t_2}(k_2)\;
\biggr)\;.
\ea
Since $\Delta{\cal M}$ is proportional to the neutrino
mass, one might expect the square of the Majorana amplitude, $|{\cal
  M}_M|^2 = |{\cal M}_D + \Delta{\cal M}|^2$,  to be equal to
the squared Dirac amplitude plus a term proportional to $m_{\nu}$
plus terms proportional to higher powers of $m_{\nu}$.
The term proportional to a single power of $m_{\nu}$, however,
vanishes since the ``interference term'' yields
\ba
{\cal M}_D^*\Delta{\cal M}+\Delta{\cal M}^*{\cal  M}_D\; = \; O(m_\nu^2)\;.
\ea

Summing over the spins of the final fermions and averaging over the
spin of the initial electron I obtain the exact final result of the squared
$\pi^\pm e^-\to\nu_\mu\mu^\pm e^-$ amplitude for the Majorana case:
\ba
\overline{|{\cal M}_{M}^{\mu^\pm}|^2}=
8\;|A|^2&\biggl\{&\;\left[(g_V^e)^2+(g_A^e)^2\right]\;\Bigl\{\;(q^2-m_\nu^2)
(q^2-m_{\mu}^2)\;
\left[\;(p_1k_2)(p_2k_3)+(p_1k_3)(p_2k_2)\;\right]\nn\\
&+&2m_{\mu}^2\;\left[\;(q^2+m_\nu^2)+
(1+\frac{m_{\nu}^2}{m_{\mu}^2})(k_3q)\;\right]\;
\left[\;(p_1k_2)(p_2q)+(p_1q)(p_2k_2)\;\right]\;\Bigr\}\nn\\
&\pm&2g_V^eg_A^e\;\Bigl\{\;(q^2+m_\nu^2)
(q^2-m_{\mu}^2)\;
\left[\;(p_1k_3)(p_2k_2)-(p_1k_2)(p_2k_3)\;\right]\nn\\
&+&2m_{\mu}^2\;\left[\;(q^2-m_\nu^2)+
(1-\frac{m_{\nu}^2}{m_{\mu}^2})(k_3q)\;\right]\;
\left[\;(p_1q)(p_2k_2)-(p_1k_2)(p_2q)\;\right]\;\Bigr\}\nn\\
&-&\left[(g_V^e)^2-(g_A^e)^2\right]\;m_e^2\;\biggl\{\;(q^2-m_\nu^2)
(q^2-m_{\mu}^2)\;(k_2k_3)\nn\\
&+&2m_{\mu}^2\;\left[\;(q^2+m_\nu^2)+
(1+\frac{m_{\nu}^2}{m_{\mu}^2})(k_3q)\;\right]\;
(k_2q)\;\biggr\}\nn\\
&\pm&2\;m_{\nu}^2\;\biggl\{\left[(g_V^e)^2+(g_A^e)^2\right]\;
\;\Bigl[\;(q^2+m_{\mu}^2)\;(p_1p_2)(k_3q)
+2m_{\mu}^2\; q^2(p_1p_2)\Bigr]\nn\\
&\pm&2\,g_V^eg_A^e\; (q^2-m_{\mu}^2)\;\Bigl[\;(p_1k_3)(p_2q)
-(p_1q)(p_2k_3)\Bigr]\nn\\
&-&2\left[(g_V^e)^2-(g_A^e)^2\right]
\;m_e^2\;\;\Bigl[\;(q^2+m_{\mu}^2)\;(k_3q)
+2m_{\mu}^2\; q^2\Bigr]\;\biggr\}\biggr\}\;.
\ea
The squared amplitude for the Dirac case,
$\overline{|{\cal M}_{D}^{\mu^\pm}|^2}$, is recovered
when setting $m_{\nu}=0$ in the above result for $\overline{|{\cal
    M}_{M}^{\mu^\pm}|^2}$.  
For all of the allowed phase space, including $q^2=0$, the result
for massive Majorana neutrinos can be expressed as
\ba
\overline{|{\cal M}_{M}^{\mu^\pm}|^2}=
\overline{|{\cal M}_{D}^{\mu^\pm}|^2}+
O\left(\frac{m_{\nu}^2}{m_{\mu}^2}\right)\;.
\ea
In the limit of vanishing neutrino mass the Majorana cross section
clearly approaches the Dirac cross section: A light Majorana neutrino
produced via the CC process $\pi^+\to \mu^++\nu_{\mu}$ behaves
approximately like a Dirac neutrino because of the left-handedness of the
CC vertex.  Similarly, a Majorana neutrino produced via $\pi^-\to
\mu^-+\nu_{\mu}$ behaves approximately like a Dirac anti-neutrino.
Although the Majorana $\nu\nu Z$ vertex is pure axial-vector, the
polarization of the incoming muon neutrino leads to a suppression of
the right-handed contribution to the cross section. The above
results illustrate point e) of the introduction.

In
the case of Dirac neutrinos one can in good approximation calculate 
neutrino cross
sections by only considering the sub-reaction (\ref{subreact}).  In the
case of  Majorana neutrinos this, however, yields a wrong result.
Instead of calculating the complete $2\to 3$ process as done above,
one can evaluate the sub-reaction (\ref{subreact})
taking into account a ``state preparation factor'' \cite{shrock} for
the incoming Majorana neutrino (see Appendix A). This gives a good 
approximation, but
neglects effects of $O(m_\nu)$.

Taking into account the upper bound for the muon neutrino mass,
$m_{\nu_{\mu}} < 0.17\;\mbox{MeV}$ at $90\%$ confidence
level\cite{partdata}, the difference between Dirac and Majorana cross
sections is too small to be detected by the CHARM II experiment.  

%
%
\section{Summary and Comments}

Typical neutrino experiments such as CHARM II are based on the fact
that the incident neutrinos are created via purely left-handed CC
processes.      
This makes it very
difficult to experimentally distinguish 
between Majorana and Dirac neutrinos: The Majorana cross section
smoothly approaches the Dirac result as $m_\nu\to0$.  Differences in
the two cross sections are of $O(m_\nu/M)$, where $M$ is a
typical energy scale of the process.
So far such differences are below the detection
limits of present-day experiments.  A recent claim \cite{plaga} that
the neutrino events observed by the CHARM II experiment give strong
evidence for the absence of Majorana
neutrinos is incorrect. This was already noticed by Kayser\cite{kayser-c}.
The main error is the neglection of the fact that the incident
neutrinos are produced via a left-handed interaction. The 
calculation in Sec.~III and its generalization in Appendix~B show this
explicitly.  

The effect of the left-handed production mechanism can {\it approximately} be
formulated by the introduction of a state preparation factor
\cite{shrock}. This approximation is valid for small 
neutrino masses; see (A.15).  Not taking into account the left-handed
production mechanism, that is, neglecting the state preparation
factor, one obtains incorrect Majorana amplitudes.  
In the case of Dirac neutrinos, the state preparation factor either
leaves the amplitude approximately unchanged 
or leads to approximately non-interacting (``sterile'') Dirac neutrinos.
In the case of massless chiral neutrinos there is no distinction
between Majorana and Dirac particles. Hence their cross sections are
identical \cite{shrock} as is pointed out in \cite{hanne}.

To become more sensitive to the Majorana or Dirac nature of neutrinos,
it would be ideal to avoid the presence of the ``state preparation
factor''.  
An obvious example
for such a process is the production of neutrino pairs.
If such neutrinos are massive Majorana particles, their transverse
polarization is {\it not} suppressed by the smallness of their mass. 
In contrast, Dirac neutrinos are almost eigenstates of helicity.   
For the NC process $e^+e^-\to Z\to\nu_f\nu^{\!\!\!\!\!\!(-)}_f$ I have
pointed out the existence of Majorana neutrino cross sections which
are {\it not} smooth as $m_\nu\to0$: Recall the forward-backward
asymmetries of (2.4) and (2.5).  Such observables are ideal candidates
for distinguishing Majorana and Dirac neutrinos. The existence of this
possibility was not recognized in the discussion given in \cite{shrock}.
The challenge, of course, is to identify experiments with high enough
luminosity to collect enough events.

Another possibility is to change the ``state preparation
factor'' for CC produced neutrinos.
Massive Dirac neutrinos have a radiatively induced magnetic 
moment (see e.g.~\cite{kayser-el}).
Massive Majorana neutrinos possibly have a magnetic
transition moment (see e.g.~\cite{valle}), connecting two different mass
eigenstates with 
opposite chiral preparation.  Though these moments are expected to be very
small, the application of extremely strong magnetic fields can lead to
a state transition of the neutrino: The neutrino can lose its memory
of its chiral production.  Effectively, the ``state preparation
factor'' of neutrinos is altered via magnetic fields. Taking into
account the smallness of the neutrino mass, this is approximately the
same as flipping the neutrino's helicity. For
example, Dirac neutrinos produced in the sun could be turned
non-interacting (``sterile'') when passing through the strong solar
magnetic field. This is considered as a possible solution to the solar
neutrino problem (see e.g.~\cite{moha}). On the other hand, Majorana 
neutrinos which have traveled
through the same magnetic field would still be interacting via
left-handed currents. In fact, CC produced Majorana neutrinos which
would behave as if they were Dirac neutrinos (anti-neutrinos) are
changed into a state such that they then behave as Dirac
anti-neutrinos (neutrinos).
%
\acknowledgements
I am very grateful to S.~Herrlich, F.~Jegerlehner, K.~Riesselmann
and A.~Vicini for many interesting and illuminating discussions as
well as for carefully  
reading the manuscript. Especially K.~Riesselmann did a lot of effort
to support my work.
%
\appendix
\section{Basic properties of neutrinos}
\setcounter{equation}{0}
\renewcommand{\theequation}{A.\arabic{equation}}
The quantized wave function for a free Majorana field may 
be written as
\be   
\nu(x) = \int\,\frac{d^3p}{(2\pi)^3 2p_0}\;\sum_{s=\pm}\;
\left[\;  f_s(\vec p)\;u_s(\vec p)\; e^{-ipx} + 
\lambda\;f^{\dagger}_s(\vec p)\;v_s(\vec p)\;e^{ipx}\;\right]\;,
\ee 
$f_s(\vec{p})$ and $f^{\dagger}_s(\vec{p})$ being the annihilator and
creator of a free one particle state with helicity $s$,
and $\lambda$ being an arbitrary phase factor (see e.g.~\cite{moha,gibrat}).
In general an amplitude for a certain process can be derived from the
interaction Lagrangian by S-matrix expansion, using Wick's theorem. 
In processes with massive Majorana particles, compared to processes
with Dirac neutrinos, additional Wick contractions of field 
operators can appear, due to the fact that there are only two
possible states: a neutrino with helicity~$+$ and a neutrino with
helicity~$-$ 
which are created by $f^{\dagger}_+$ and $f^{\dagger}_-$, respectively.
The two helicity states are connected by a Lorentz transformation.
A detailed discussion on how the matrix element for massive Majorana neutrinos
can be calculated using the charge conjugation property of the 
free Majorana field is given in \cite{denner}.

The neutral current of a massive
Majorana neutrino,
\ba
j_{\lambda}(x)=\bar{\nu}(x)\;\gamma_{\lambda}\;(g_V^{\nu}-g_A^{\nu}\gamma_5)\;
\nu(x)\;,
\ea
is pure axial-vector for arbitrary vector and
axial-vector couplings $g_V^\nu$ and $g_A^\nu$ in the Lagrangian.
Using the charge conjugation Dirac matrix $C$ and taking the
neutral current to be
normal-ordered\footnote{For the calculation of S matrix elements, only
time-ordered Green's functions are relevant. It can be shown that 
therefore only the normal-ordered part of the neutral current
contributes.}, one can write:
\ba
\label{axialcurr}
j_{\lambda}(x)\;
&=&
\;\bar{\nu}(x)\;\gamma_{\lambda}\;(g_V^{\nu}-g_A^{\nu}\gamma_5)\;
\nu(x)\;=\;-\nu^T(x)\;C^{-1}C\;(g_V^{\nu}-g_A^{\nu}\gamma_5)\;
\gamma_{\lambda}^T\;C^{-1}C\;\bar{\nu}^T(x)\;\nn\\
&=&\;\bar{\nu}(x)\;C\;(g_V^{\nu}-g_A^{\nu}\gamma_5)\;
    \gamma_{\lambda}^T\;C^{-1}\;\nu(x)
\;=\;-\bar{\nu}(x)\;\gamma_{\lambda}\;(g_V^{\nu}+g_A^{\nu}
    \gamma_5)\;\nu(x)\;,
\ea
where the following identities have been used:
\ba
C\bar{\nu}^T&=&\lambda^*\nu\;,\quad C^T=C^{-1}=-C\;,\quad 
C\gamma_{\lambda}^TC^{-1}=-\gamma_{\lambda}
\quad\mbox{and}\quad C\gamma_5\gamma_{\lambda}^TC^{-1}=
\gamma_{\lambda}\gamma_5\;\;.
\ea
From (\ref{axialcurr}) the pure axial-vector nature of the neutral
current is apparent:
\ba
j_{\lambda}(x)\;=-g_A^{\nu}\;\bar{\nu}(x)\;\gamma_{\lambda}\;\gamma_5\;
\nu(x)\;\;.
\ea

I now prove point a) of the introduction. In the massless case 
the Dirac equation only has two linear independent solutions:
\ba
&&u_{L,R}\;\propto\; v_{R,L}\;,
\ea
where 
\ba
u_{L,R}=P_{L,R}\;u\;, \qquad v_{R,L}=P_{L,R}\;v\;,
\ea 
and $P_{L,R}$ are the chiral projectors as defined in (3.6). In addition,
chirality and helicity are the same for massless neutrinos:
\ba
\label{same}
&&u_{L,R}=u_{-,+}\;,\qquad v_{L,R}=v_{-,+}\;.
\ea
If only the left-handed part\footnote{Of course the same discussion 
could be made for the right-handed part.} of the massless Majorana field 
interacts, only the left-handed chiral projection of (A.1) is
relevant. Because of (\ref{same}) I immediately obtain
\be   
\nu_L(x) = \frac{1}{2}(1-\gamma_5)\;\nu(x) =
\int\,\frac{d^3p}{(2\pi)^3 2p_0}\;
\left[\;  f_-(\vec p)\;u_-(\vec p)\; e^{-ipx} + 
\lambda\;f^{\dagger}_+(\vec p)\;v_+(\vec p)\;e^{ipx}\; \right]\;.
\ee
Since $f_-$ and $f_+$ are independent operators obeying the
anti-commutation relations of the Dirac algebra $\nu_L(x)$ has the
well-known form of the quantized field of a left-handed massless Dirac
neutrino.\footnote{The phase factor $\lambda$ can be absorbed by
  redefining $f_+$.} Therefore a massless Majorana neutrino behaves in
the same way as a massless Dirac neutrino if only left-handed (or only
right-handed) interactions are present.  End of proof. Please notice
that the two different helicity states corresponding to $f_-$ and
$f_+$ are the same in all Lorentz frames because the massless neutrino
is traveling with the speed of light. This is not the case as soon as
the neutrino has a non-zero mass. In particular, there is no smooth
limit for restoring the Lorentz invariance of the helicity as
$m_\nu\to0$ or $\beta\to1$. 
There is an additional fundamental difference between massless
Majorana neutrinos 
on one hand and light or relativistic Majorana neutrinos on the other
hand: Only in the massless case chirality is a good quantum number,
being then identical to helicity.

Next I show that the neutral current for a {\it massless} Majorana
neutrino is chiral. For both Dirac and Majorana neutrinos the neutral
current of (A.2) can be rewritten as
\ba
\label{chircurr}
j_{\lambda}(x)\;=\;
(g_V^{\nu}+g_A^{\nu})\;\bar\nu_L(x)\;\gamma_{\lambda}\;\nu_L(x)\;+\;
(g_V^{\nu}-g_A^{\nu})\;\bar\nu_R(x)\;\gamma_{\lambda}\;\nu_R(x)\;.
\ea
Because massless chiral Majorana fields satisfy
$C \bar\nu_{L,R}^T(x)=\lambda^*\,\nu_{R,L}(x)\;$, 
the following identity is obtained:
\ba
\bar\nu_L(x)\;\gamma_{\lambda}\;\nu_L(x)\;=
\;-\bar\nu_R(x)\;\gamma_{\lambda}\;\nu_R(x)\;\;
\ea
Inserting this result in (\ref{chircurr}) I find that even for
arbitrary vector and axial-vector couplings  
the neutral current for massless Majorana neutrinos
is chiral: 
\ba
\label{chircurr2}
j_{\lambda}(x)\;=\;2g_A^{\nu}\;\bar\nu_L(x)\;\gamma_{\lambda}\;\nu_L(x)
\;=\;-2g_A^{\nu}\;\bar\nu_R(x)\;\gamma_{\lambda}\;\nu_R(x)\;\;.
\ea
As was proven above, the massless chiral Majorana field $\nu_L$ 
($\nu_R$) can equivalently be treated as a massless chiral
Dirac field if only $\nu_L$ ($\nu_R$) interacts.
For Dirac neutrinos the current $j_{\lambda}$ in (\ref{chircurr}) is
only chiral if 
\ba
\label{chircurr3}
g_V^{\nu}=\pm g_A^{\nu}\;.
\ea
Comparison of (\ref{chircurr}) with (\ref{chircurr2}) shows that massless
Dirac and Majorana neutrinos are the same if and only if (\ref{chircurr3}) is valid. 

A comparison of (A.5) with (\ref{chircurr2}) confirms the statement d)
of the introduction that for Majorana neutrinos the small-mass limit
is in general not approaching the massless case: While the $\nu\nu Z$
vertex involving massive Majorana neutrinos is always axial-vector, it
is always chiral in the case of massless Majorana neutrinos.
Additionally, $g_V^{\nu}$ cannot be measured if neutrinos
are Majorana particles, disregarding whether they are massive or
massless.  

Next I consider neutrinos which are produced in a
chiral interaction such as in a CC interaction.
Because of its very small mass,
a chirally produced neutrino behaves in good 
approximation as if the initial neutrino state contains a
preparation factor $\frac{1}{2}(1\mp\gamma_5)$.
For Dirac neutrinos the state preparation factor can be left away
if neutrinos exclusively  couple left-handedly 
(or, equivalently, only right-handedly) and if terms of $O(m_\nu)$ are 
neglected.
However for Majorana neutrinos the state preparation factor plays an
important role. 
It was shown above that the NC Majorana neutrino vertex is pure axial 
vector, hence [using $j_{\lambda}(x)$ in (A.2)]: 
\ba
\langle\nu(p_f,s_f)|\;\bar{\nu}(x)\;\gamma_{\lambda}\;
(g_V^{\nu}-g_A^{\nu}\gamma_5)\;\nu(x)\;|\nu(p_i,s_i)\rangle
\;=\;-2g_A^{\nu}\;\bar u(\vec{p}_f,s_f)\;\gamma_{\lambda}\gamma_5\;
u(\vec{p}_i,s_i)\\\times\; e^{-i(p_i-p_f)x}\nn.
\ea
In contrast a calculation for a chirally prepared initial Majorana
state yields 
\ba
\langle\nu(p_f,s_f)|\;\bar{\nu}(x)\;\gamma_{\lambda}\;
(g_V^{\nu}-g_A^{\nu}\gamma_5)\;\nu(x)\;|\nu_{L,R}(p_i,s_i)\rangle
\;=\;-2g_A^{\nu}\;\bar u(\vec p_f,s_f)\;\gamma_{\lambda}P_{L,R}
\;u(\vec p_i,s_i)\\
\times\; e^{-i(p_i-p_f)x}\;+\;O(m_\nu)\;=\;
-2g_A^{\nu}\;\bar v(\vec p_i,s_i)\;\gamma_{\lambda}P_{R,L}
\;v(\vec p_f,s_f)\; e^{-i(p_i-p_f)x}\;+\;O(m_\nu)\;\;.\nn
\ea
The chirally prepared initial Majorana state $|\nu_{L,R}(p,s)\rangle$ in
(A.15) is defined as   
\ba
|\nu_{L,R}(p,s)\rangle &=& f_{L,R}^{\dagger}(p,s)|0\rangle\;,\\
f_{L,R}^{\dagger}(p,s) &=& \int d^3x\;\bar\nu(x)\;\gamma_0\;
P_{L,R}\;u(p,s)\;e^{-ipx} \nn\\
&=&\int d^3x\;\bar v(p,s)\;\gamma_0\;
P_{R,L}\; \nu(x)\;e^{-ipx}\;.
\ea
In comparison, for chirally prepared Dirac neutrinos 
(anti-neutrinos) one obtains
\ba
\langle\nu(p_f,s_f)|\;\bar{\nu}(x)\;\gamma_{\lambda}\;
(g_V^{\nu}-g_A^{\nu}\gamma_5)\;\nu(x)\;|\nu_L(p_i,s_i)\rangle
\;&=&\;(g_V^{\nu}+g_A^{\nu})\;\bar u(\vec p_f,s_f)\;\gamma_{\lambda}P_L
\;u(\vec p_i,s_i)\nn\\
&\times&\;e^{-i(p_i-p_f)x}\;+\;O(m_{\nu})\;,\\
\langle\bar\nu(p_f,s_f)|\;\bar{\nu}(x)\;\gamma_{\lambda}\;
(g_V^{\nu}-g_A^{\nu}\gamma_5)\;\nu(x)\;|\bar\nu_L(p_i,s_i)\rangle
\;&=&-(g_V^{\nu}-g_A^{\nu})\;\bar u(\vec
p_f,s_f)\;\gamma_{\lambda}P_L\;u(\vec p_i,s_i)\nn\\
&\times&\;e^{-i(p_i-p_f)x}
\;+\;O(m_{\nu})\;,\\
\langle\nu(p_f,s_f)|\;\bar{\nu}(x)\;\gamma_{\lambda}\;
(g_V^{\nu}-g_A^{\nu}\gamma_5)\;\nu(x)\;|\nu_R(p_i,s_i)\rangle
\;&=&\;(g_V^{\nu}-g_A^{\nu})\;\bar v(\vec p_i,s_i)\;\gamma_{\lambda}P_L\;
v(\vec p_f,s_f)\nn\\
&\times&\;e^{-i(p_i-p_f)x}\;+\;O(m_{\nu})\;,\\
\langle\bar\nu(p_f,s_f)|\;\bar{\nu}(x)\;\gamma_{\lambda}\;
(g_V^{\nu}-g_A^{\nu}\gamma_5)\;\nu(x)\;|\bar\nu_R(p_i,s_i)\rangle
\;&=&-(g_V^{\nu}+g_A^{\nu})\;\bar v(\vec p_i,s_i)\;\gamma_{\lambda}P_L
\;v(\vec p_f,s_f)\nn\\&\times&\;e^{-i(p_i-p_f)x}\;+\;O(m_{\nu})\;\;.
\ea
From (A.15), (A.18) and (A.21) it is apparent that for SM couplings
($g_V^{\nu}=g_A^{\nu}$) a left-handedly prepared Majorana neutrino 
behaves like a Dirac neutrino while a 
right-handedly prepared Majorana neutrino behaves like a 
Dirac anti-neutrino, up to terms of $O(m_\nu)$.
In contrast a right-handedly prepared Dirac neutrino as well as a
left-handedly prepared Dirac anti-neutrino would
be ``sterile'' in respect to left-handed NC\footnote{It
is easy to show that this is also the case for left-handed CC 
interactions.} interactions, because (A.19) and (A.20) are of $O(m_\nu)$. 
(A.18)-(A.21) furthermore show that in the case of a chiral NC
the state preparation factor may be left away for Dirac neutrinos if
terms of $O(m_\nu)$ are neglected.

\section{Cross sections for generalized couplings}
\renewcommand{\theequation}{B.\arabic{equation}}
\setcounter{equation}{0}
The case of arbitrary vector and axial-vector coupling
$g_V^{\nu}$ and  $g_A^{\nu}$ of the 
neutral neutrino current      
is considered. I discuss the restrictions 
on the general couplings $g_V^{\nu}$ and $g_A^{\nu}$ 
which can be obtained 
from the two neutral current experiments LEP I and CHARM II.

In the case of LEP I ($s\approx m_Z^2$), the generalized differential
cross sections for 
neutrino pair production for the different cases of Majorana and Dirac
neutrinos with or without a mass are
\ba
\left(\frac{d\sigma}{d\Omega}\right)_{M}^{m_{\nu}\not=0}
&=&4\;\sigma_0\;
(g_A^{\nu})^2\;\biggl\{\left[(g_V^e)^2+(g_A^e)^2\right]\;(1+\cos^2\theta)\nn\\
&&\qquad\qquad+4\left[(g_V^e)^2-(g_A^e)^2\right]\frac{m_e^2}{s}
+O\left(\frac{m_{\nu}^2}{s}\right)\biggr\}\;\;,\\
\left(\frac{d\sigma}{d\Omega}\right)_{M}^{m_{\nu}=0} &=&4\;\sigma_0\;
(g_A^{\nu})^2\;\biggl\{
\left[(g_V^e)^2+(g_A^e)^2\right](1+\cos^2\theta)
+4g_V^eg_A^e\cos\theta\nn\\
&&\qquad\qquad+4\left[(g_V^e)^2-(g_A^e)^2\right]\frac{m_e^2}{s}\biggr\}\;\;,\\
\left(\frac{d\sigma}{d\Omega}\right)_{D}^{m_{\nu}\not=0}
&=&2\;\sigma_0\;
\left[(g_V^{\nu})^2+(g_A^{\nu})^2\right]\;\biggl\{
\left[(g_V^e)^2+(g_A^e)^2\right](1+\cos^2\theta)\nn\\
& & +8\frac{g_V^eg_A^eg_V^{\nu}g_A^{\nu}}{(g_V^{\nu})^2+(g_A^{\nu})^2}
\cos\theta
+4\left[(g_V^e)^2-(g_A^e)^2\right]\frac{m_e^2}{s}+
O\left(\frac{m_{\nu}^2}{s}\right)\biggr\}\;\;,
\ea
and $({d\sigma}/{d\Omega})_{D}^{m_{\nu}=0}$ is obtained from (B.3) by
setting $m_\nu=0$. The quantity $\sigma_0$ is defined in (2.3).
At LEP I the sum of the total cross sections for neutrino pair
production has been measured indirectly and is 
(assuming three light neutrino generations)
\ba
\sigma_{\rm LEP}&=&\sum_{f=e,\mu,\tau}\sigma(e^+e^-\to
\nu_f\nu^{\!\!\!\!\!\!^{(-)}}_f)\nn\\
&=&\frac{16}{3}\;\sigma_0\;\left[(g_V^e)^2+(g_A^e)^2\right]\sum_{f=e,\mu,\tau}
\left\{ \begin{array}{r@{\quad:\quad}l} 
\left[(g_V^{\nu_f})^2+(g_A^{\nu_f})^2\right] & {\rm for\;\; Dirac,}
\\
2\;(g_A^{\nu_f})^2 & {\rm for\;\; Majorana.}
\end{array} \right. 
\ea
The factor $(g_V^e)^2+(g_A^e)^2$ can be determined 
by the measurement of the
partial width of $e^+e^-$ production at LEP. 
The calculated cross sections are consistent with the measured data
if the couplings satisfy
\ba
\sum_{f=e,\mu,\tau}\left[(g_V^{\nu_f})^2+(g_A^{\nu_f})^2\right]&=&
\frac{3}{2}\qquad\mbox{for Dirac neutrinos\hspace{3.0cm}}
\ea
or
\ba
\sum_{f=e,\mu,\tau}(g_A^{\nu_f})^2&=&\frac{3}{4}\qquad
\mbox{for Majorana neutrinos ($g_V^{\nu_f}$ being arbitrary)}\;\;.
\ea

The amplitudes of Sec.~III for NC $\nu_\mu e^-$ scattering need
also to be modified for arbitrary couplings $g_V^{\nu}$ and $g_A^{\nu}$.
The new amplitudes $\widetilde{{\cal M}}_{D}$ and 
$\widetilde{{\cal M}}_{M}$ can be written as combinations of
the amplitudes ${\cal M}_D$ and $\Delta{\cal M}$ defined in (3.2), (3.4), 
(3.8) and (3.9):
\ba
\widetilde{{\cal M}}_M^{m_{\nu}\not=0}\;&=&\;2g_A^{\nu}\cdot
({\cal M}_D+\Delta{\cal M})\;,\\
\widetilde{{\cal M}}_M^{m_{\nu}= 0}\;&=&\;2g_A^{\nu}\cdot {\cal M}_D\;,\\
\widetilde{{\cal M}}_D^{m_{\nu}\not=0}\;&=&\;(g_V^{\nu}+g_A^{\nu})\cdot{\cal M}_D
+ (g_V^{\nu}-g_A^{\nu})\cdot\Delta{\cal M}\;,\\
\widetilde{{\cal M}}_D^{m_{\nu}=0
    }\;&=&\;(g_V^{\nu}+g_A^{\nu})\cdot{\cal M}_D\;.
\ea
Summing over final
spins and averaging over the initial electron spin
the squared matrix elements are
\ba
\overline{|\widetilde{{\cal M}}_M^{m_{\nu}\not=0}|^2}\;&=&
\;4(g_A^{\nu})^2\;\overline{|{\cal M}_M|^2}
\;=\;4(g_A^{\nu})^2\;\overline{|{\cal M}_D|^2}+
O\left(\frac{m_{\nu}^2}{m_{\mu}^2}\right)\;\;,\\
\overline{|\widetilde{{\cal M}}_M^{m_{\nu}= 0}|^2}\;&=&
\;4(g_A^{\nu})^2\;\overline{|{\cal M}_D|^2}\;\;.\\
\overline{|\widetilde{{\cal M}}_D^{m_{\nu}\not=0}|^2}\;
&=&\;(g_V^{\nu}+g_A^{\nu})^2\;
\overline{|{\cal M}_D|^2}
+\left[(g_V^{\nu})^2-(g_A^{\nu})^2\right]\;
(\overline{{\cal M}_D^*\Delta{\cal M}}+\overline{\Delta{\cal M}^*
{\cal  M}_D})\nn\\
&&\;+
(g_V^{\nu}-g_A^{\nu})^2\;\overline{|\Delta{\cal M}|^2}\nn\\
&=&\;(g_V^{\nu}+g_A^{\nu})^2\;\overline{|{\cal M}_D|^2}+
O\left(\frac{m_{\nu}^2}{m_{\mu}^2}\right)\;\;,\\
\overline{|\widetilde{{\cal M}}_D^{m_{\nu}=0}|^2}\;
&=&\;(g_V^{\nu}+g_A^{\nu})^2\;
\overline{|{\cal M}_D|^2}\;\;.
\ea
The CHARM II collaboration has also measured $\nu_ee^-$ scattering. 
This channel includes 
a CC contribution.  

Again the cross sections can be used
to determine the couplings
$g_V^{\nu_f}$ and $g_A^{\nu_f}$ ($f=e,\mu$).
The experimental data are consistent with the conditions
\ba
(g_V^{\nu_f}+g_A^{\nu_f})^2&=&1\qquad\mbox{for Dirac neutrinos,}
\qquad\\    
(g_A^{\nu_f})^2&=&\frac{1}{4}\qquad
\mbox{for Majorana neutrinos ($g_V^{\nu_f}$ being arbitrary)}\;\;.
\ea

Combining the LEP I results in (B.5) (where the existence of three
light neutrino generations has been assumed) and the CHARM II results
in (B.15) (assuming that
this result is also true for tau Dirac neutrinos) 
the numerical values of $g_{V,A}^{\nu_f}$ are determined for Dirac neutrinos:
\ba
g_V^{\nu_f}=g_A^{\nu_f}=\frac{1}{2},\quad f=e,\mu,\tau\;\;.
\ea
Under the above assumptions, the neutral Dirac neutrino current 
has to be left-handed 
and flavour universality is obtained. These results are in complete
agreement with the standard model. 

Correspondingly, the LEP I and CHARM II results can also be combined
assuming Majorana neutrinos.
From (B.6) (again assuming three light neutrino generations) and
(B.16) (without any assumptions regarding the tau Majorana neutrino)
it follows that 
\ba
g_A^{\nu_f}=\frac{1}{2},\quad
g_V^{\nu_f}\;\mbox{being arbitrary},\quad f=e,\mu,\tau\;.
\ea

Under the assumption of SM couplings and flavour universality for all
light neutrino flavours, the LEP I measurements have determined the
number of light (Dirac or Majorana) neutrinos to be three. 
Clearly, the CHARM II experiment has confirmed
the assumption of flavour universality for two neutrino flavours,
namely for the electron and muon family, independent of the fact whether
neutrinos are Dirac or Majorana particles.  Presently, there is no
experimental proof that the tau neutrino has the same couplings as the
electron neutrino and muon neutrino.

\newpage
\begin{center}
{\large FIGURE CAPTIONS}
\end{center}
\begin{itemize}
\item[1.] The Feynman diagram of the process
  $e^+e^-\to Z\to\nu\nu$. The neutrinos can be either Dirac or
  Majorana particles.
\item[2.] The Feynman diagram of the $2\to3$ process $\pi^+
  e^-\to\nu_\mu\mu^+ e^-$.  The neutrino can be either a Dirac or
  Majorana particle.  
\end{itemize}
\newpage    
\begin{figure}[H]
\begin{center}
\hspace{2cm}\mbox{\epsfysize 7cm \epsffile{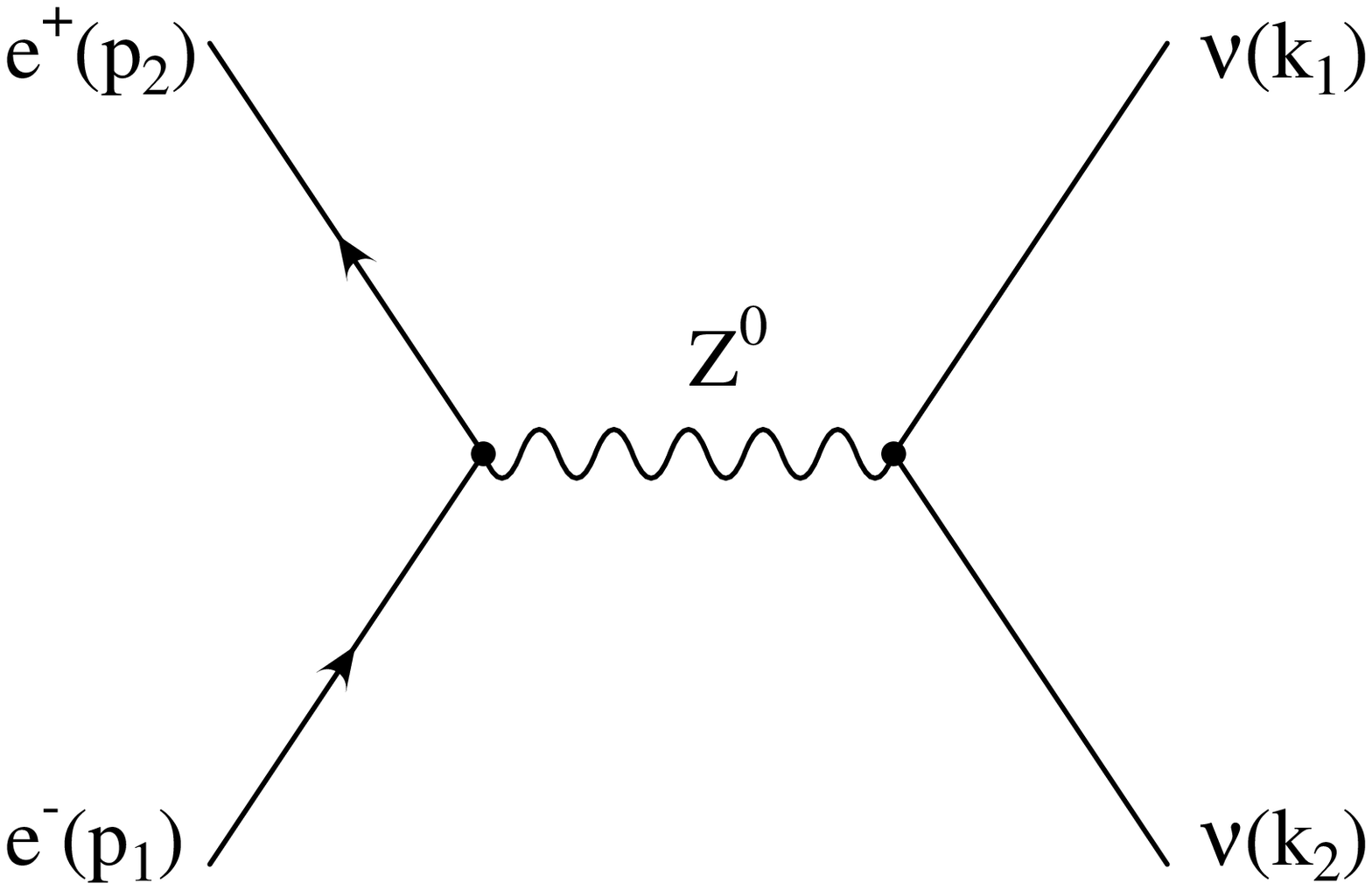}}
\vspace{1.5cm}
\caption{}
\end{center}
\end{figure}
\vspace{1cm}
\begin{figure}[H]
\begin{center}
\vspace{-2cm}
\mbox{\epsfysize 9cm \epsfxsize 12cm \epsffile{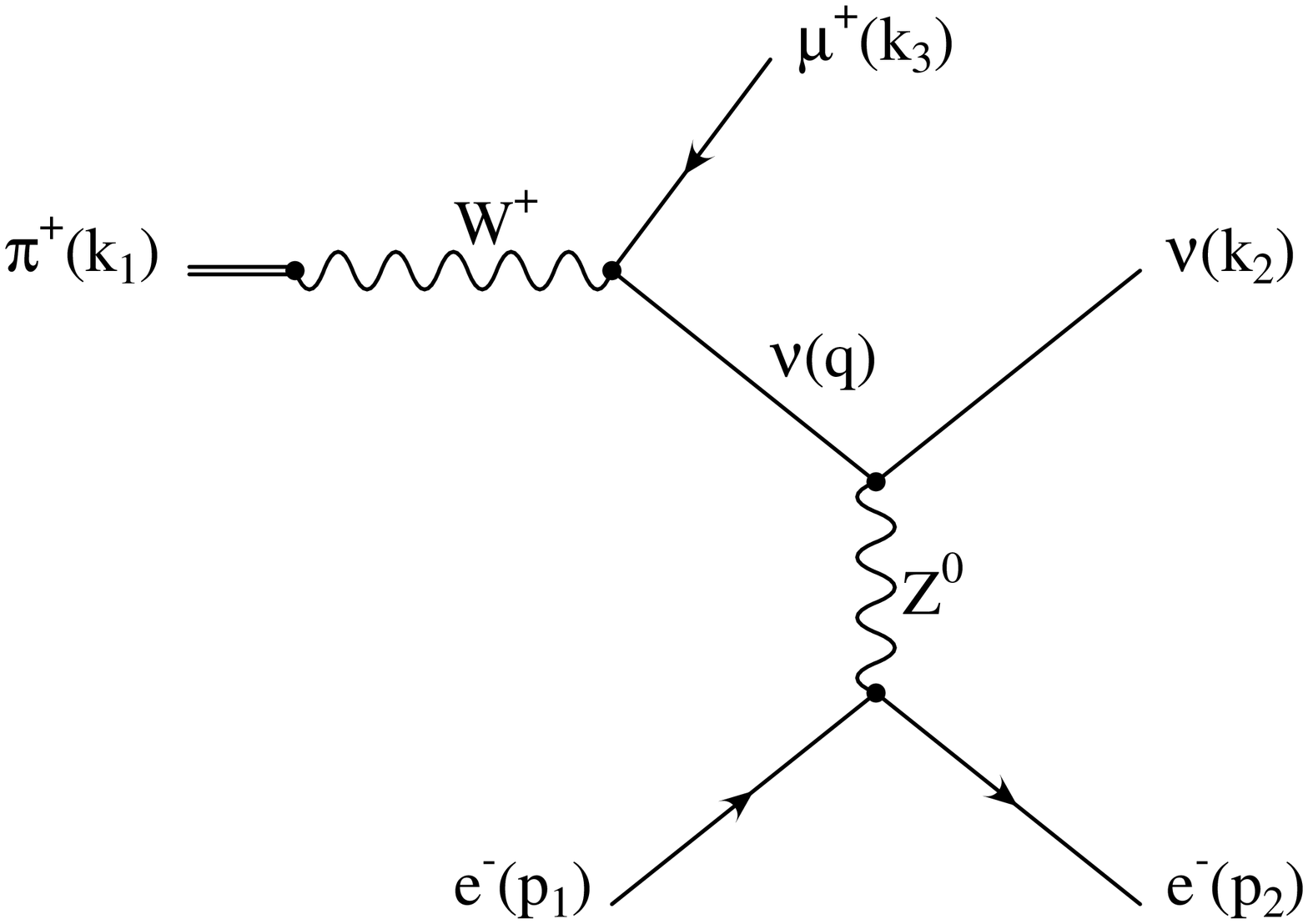}}
\vspace{1.7cm}
\caption{}
\end{center}
\end{figure}
\end{document}